\documentclass[twocolumn,showpacs,amssymb,nobibnotes,aps,floats,psfig,prb]{revtex4-1}
\usepackage{textcomp,amssymb,graphicx,epsfig}
\usepackage{hyperref}
\usepackage{amsbsy}
\usepackage{amsmath,amsfonts,color}
\usepackage{float}
\date{}

\begin{document}
\title{A study of cooperative breathing-mode in molecular chains}
\author{Ravindra Pankaj$^1$}
\author{Sudhakar Yarlagadda$^{1,2}$}
\affiliation{${^1}$TCMP Div., Saha Institute of Nuclear Physics,
Kolkata, India}
\affiliation{${^2}$Cavendish Lab, Univ. of Cambridge, Cambridge, UK}
\date{\today}
\begin{abstract}
Using a controlled analytic non-perturbative treatment, that accounts for the quantum 
nature of the phonons,
we derive a model that generically describes cooperative breathing-mode at
strong electron-phonon interaction in one-band one-dimensional
 systems. The effective model
involves a {\em next-nearest-neighbor} hopping (that dominates over the nearest-neighbor
hopping at strong coupling)
and a nearest-neighbor repulsion that is significantly enhanced due to incompatibility of neighboring
dilations/compressions.
 At non-half filling, upon tuning the electron-phonon coupling,  the system
 undergoes a period-doubling second-order quantum phase transition from a Luttinger liquid
to a {\em conducting commensurate}
charge-density-wave state: a phenomenon absent in both the Holstein model and the t-V model.
 Using fidelity to study the nature of the quantum phase transition, we find that the fidelity susceptibility 
shows a superextensive power law divergence as well as 
 a remarkable scaling behavior: both together establish a second-order transition. 
\end{abstract}
\maketitle
\section{Introduction}
Perovskite materials are quite ubiquitous and exhibit a variety of interesting
and intriguing  
 phenomena such as
 superconductivity or charge ordering (or their co-existence),
colossal magnetoresistance, ferroelectricity, spin-dependent-transport,
and the interplay among magnetic, structural, and transport
 properties \cite{tvr1,hotta,khomskii1}.
Many oxides, that have the formula $ABO_3$, assume  a perovskite structure
where two adjacent $BO_6$ octahedra  share an oxygen which leads to cooperative
octahedral distortions.
Simple systems that manifest such cooperative electron-phonon phenomena are
the barium bismuthates ($BaBiO_3$).
Here, only the 6s electrons are involved in transport and these electrons
produce only a single normal mode distortion, namely, the breathing-mode.
In pure $BaBiO_3$, the $BO_6$ octahedra alternately dilate and contract
with $Bi-O$ bonds of adjacent octahedra differing by about $10\%$ 
which is indicative of strong electron-phonon interaction (EPI) \cite{tvr1}.
Thus the relevant physics is dominated by a one-band three-dimensional
cooperative breathing-mode (CBM).

There is also compelling evidence of strong EPI
in manganites (from
 extended X-ray absorption fine structure\cite{bianc} and
 pulsed neutron diffraction\cite{louca} measurements) and
in cuprates (through angle-resolved photoemission 
spectroscopy\cite{damascelli}).

In copper oxides, as pointed out in Refs.~\onlinecite{sawatzky,berciu}, the dynamics
of the Zhang-Rice singlet \cite{zhang} can be described by a one-band
system with orbitals centered on copper sites. Furthermore, 
the onsite energy is modulated by the movement of oxygen
closer or further from the neighboring copper ion. Thus the breathing-mode
is relevant to describe the linear modulation of the onsite energy.
Consequently the copper-oxide planes represent a one-band two-dimensional
CBM system.

In the context of the two-band Jahn-Teller manganite systems as well, when C-type
antiferromagnetism manifests [as in $La_{1-x}Sr_{x}MnO_3$ for $0.65 \le x \le 0.9$ \cite{dabrowski}],
the $d_{z^2}$ orbitals participate in the C-chain ordering. A ferromagnetic C-chain  can be looked upon as
a one-band (i.e., $d_{z^2}$ orbital  band) and one-dimensional (1D) CBM 
system that is however Jahn-Teller coupled to
neighboring C-chains whose spin alignment is antiparallel.

\begin{figure}[t]
\includegraphics[width=3.0in]{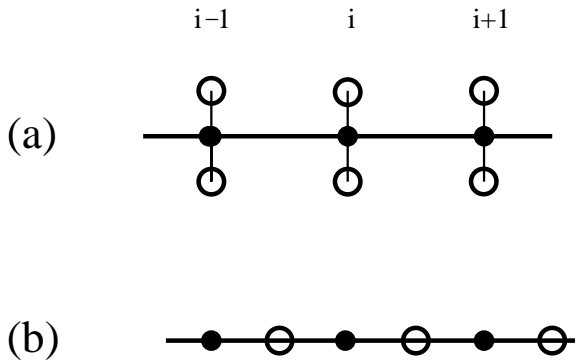}
\noindent\caption[]{Molecular chains with  $d_{z^2}$ orbital hopping sites
(filled circles) and oxygen sites (empty circles) in
(a) Holstein model,
(b) one-band CBM system.}
\label{fig:chain}
\end{figure}

Understanding the CBM phenomena, in  systems
such as the bismuthates, the cuprates, and the manganites is
still an open question. The main purpose of this paper
is to study the CBM physics in the simpler case of a 
one-band 1D system by taking account of the {\em quantum phonons} [see Fig.~\ref{fig:chain}(b)].
In fact, a controlled analytic treatment of the many-polaron effects
produced by quantum phonons
in a one-band 1D Holstein model [see Fig.~\ref{fig:chain}(a)]
\cite{holstein}
(which is a simpler non-cooperative EPI system) has 
been reported not long ago\cite{sdadys,sdys}.
However, definite progress has been made in numerically treating
 the Holstein model at half-filling (by employing a variety of
 techniques) \cite{fehske2011,fradkin2,capone,zheng1,perroni,hamer}
and, to a limited extent, away from half-filling \cite{fehske3}.

Owing to its cooperative nature, the EPI leads to non-local
distortion effects which can change the very nature of long range order.
While a weak interaction is amenable to a Migdal-type  of perturbative
treatment, the strong interaction (even for a one-band system)
necessitates a non-perturbative approach \cite{sdadys}.
As a step towards modeling CBM distortions in
real systems (such as the bismuthates, the cuprates, and the manganites),
the present work builds up on our previous work on the Holstein model \cite{sdadys}
to obtain the effective Hamiltonian for
a one-band 1D CBM system \cite{alex3,alex4,alex5}.
Upon inclusion of cooperative effects
in the strong EPI, we show that
the system changes its dominant transport mechanism from
one of nearest-neighbor (NN) hopping to that of next-nearest-neighbor (NNN) hopping
while the effective NN electron-electron
 interaction  becomes significantly more repulsive due to incompatibility of NN
breathing mode distortions.
Away from half-filling in rings with even number of sites 
(while the Holstein system without cooperative effects
remains a Luttinger liquid at all interaction strengths),
our  model (at strong interaction), 
produces
a commensurate charge-density-wave (CDW) state which is surprisingly conducting
and whose period is independent of density. Furthermore, using scaling
of the fidelity susceptibility (FS), we demonstrate that the CDW transition
is a second-order quantum phase transition (QPT).

The paper is organized as follows. We derive an effective polaronic Hamiltonian
 starting from a CBM model in Sec.~II. Next, we present the relevant formulae
for the density-density correlation function and the structure factor in Sec.~III.
We then analyze the strong-coupling  limiting case of our CBM model in Sec.~IV.
The nature of the QPT and the long range order
in our CBM model are discussed in Sec.~V. Finally, we close
in Sec.~VI with our conclusions.
\section{Effective Polaronic Hamiltonian}
To bring out the essential physics, we begin with a 1D model of spinless electrons
 hopping in a one-band system of $d_{z^2}$ orbitals which are coupled to 
the oxygens in between, via CBM as shown in Fig.~\ref{fig:chain}(b) \cite{sawatzky}.
The Hamiltonian is expressed as
 $H = H_t + H_{ep} + H_l$
where the hopping term $H_t$, using standard notation, is given by
\begin{eqnarray}
H_t = 
- t \sum_{j} ( c^{\dagger}_{j} c_{j+1} + {\rm H.c.}) , 
\label{eq:Ht}
\end{eqnarray}
with $c_j$ ($c^{\dagger}_{j}$) being the destruction (creation)
 operator of an electron in a $d^j_{z^2}$ orbital (at site $j$).
 The EPI term $H_{ep}$ is expressed as
\begin{eqnarray}
H_{ep} = 
- g \omega_0 \sqrt{2 M \omega_0} 
 \sum_{j}  n_j
 q_j , 
\end{eqnarray}
where $g$ is the electron-phonon coupling (EPC),
$n_i = c^{\dagger}_{i} c_{i} $,
$q_i = u_i - u_{i-1}$ represents the expansion of the oxygens around the
$d^i_{z^2}$ orbital, and
 the right-hand-side (RHS) oxygen displacement 
$u_i =(a^{\dagger}_i + a_i)/{\sqrt{2 M \omega_0}}$. Furthermore,
 the lattice term 
$H_l$ representing simple harmonic oscillators is 
of the form
\begin{eqnarray}
H_{l} = 
\frac{K}{2} \sum_{j} u_j^2   
+ \frac{1}{2M} \sum_{j} p_j^2 =  
 \omega_0 \sum_{j} a^{\dagger}_{j} a_{j} .
\end{eqnarray}
 The main difference between the Holstein model and the above
cooperative Hamiltonian is that in the Holstein model  electrons at
different sites are
coupled to  different on-site molecular distortions
whereas in the present system
 the electrons on adjacent sites are coupled
to the displacement of the same in-between oxygen. 
Thus in our system, to produce an effective polaronic Hamiltonian,
 we need to devise a modification of the usual Lang-Firsov
transformation \cite{LF} so as to take into account the 
cooperative nature of the
 distortions. To meet this end, we used the following canonical
transformation
 $\tilde{H} = \exp(S) H \exp(-S)$
where
$S$ now contains the {\em difference in densities on adjacent sites}
\begin{eqnarray}
S = g \sum_j ( a_{j} - a^{\dagger}_{j})
(n_j - n_{j+1}) .
\end{eqnarray}
Then, one obtains
 $\tilde{H}  = H_0 + H_1$ where
\begin{eqnarray}
H_0 =&& 
 \omega_0 \sum_{j} a^{\dagger}_{j} a_{j} -
2 g^2 \omega_0
\sum_{j} n_j 
+ 2 g^2 \omega_0 \sum_{j} n_j n_{j+1} 
\nonumber \\
&& - t e^{-3 g^2} \sum_{j} ( c^{\dagger}_{j} c_{j+1} + {\rm H.c.}) , 
\end{eqnarray}
and 
\begin{eqnarray*}
H_1 = \sum_{j}H_{1j} =
- t e^{-3 g^2}
 \sum_{j} [ c^{\dagger}_{j} c_{j+1}
\{
{\cal{T}}_{-}^{j \dagger} {\cal{T}}^{j}_{+}
-1 \} + {\rm H.c.}] , 
\end{eqnarray*}
with
 ${\cal{T}}^{j}_{\pm} = \exp[\pm g( 2 a_{j} - a_{j-1} - a_{j+1})]$.
{\it On account of the cooperative nature of the EPI
we obtain an additional term 
$ 2 g^2 \omega_0 \sum_{j} n_j n_{j+1}$} 
involving NN repulsion in $H_0$
 and the perturbation $H_1$ now involves
phonons at three sites as opposed to phonons at only two sites as in
 the non-cooperative case.
We consider the case $ t\exp[-3 g^2] << \omega_0$ and 
perform second-order perturbation theory similar
to that in Refs.~\onlinecite{sdadys, sahinur}.
The eigenstates of $H_0$ relevant for perturbation theory
are  $|n,m \rangle \equiv |n \rangle_{el} \otimes
 |m\rangle_{ph}$ where NN occupied electronic
states are projected out
and
 $|0,0\rangle$ is the ground state (GS) with no
phonons. The corresponding eigenenergies
are  $E_{n,m}=E_{n}^{el}+E_{m}^{ph}$.  
The treatment to perform second-order perturbation theory is
an extension of the method followed in Refs.~\onlinecite{sdadys, sahinur}
and yields the following effective Hamiltonian for polarons:
\begin{eqnarray}
H^{(2)} = \sum_{i,j}
\sum_{m} \frac{\langle 0|_{ph} H_{1i} |m\rangle_{ph}
 \langle m|_{ph} H_{1j} |0\rangle_{ph}}
{E_{0}^{ph} - E_{m}^{ph}} .
\label{H^2}
 \end{eqnarray}   
{Here (as shown by using Schrieffer-Wolff transformation
in appendix A of Refs. \onlinecite{sahinur,sahinur_arxiv}),}
it must be mentioned that, when   $ t\exp[-3 g^2] << \omega_0$,
$H_0 + H^{(2)}$ represents the exact Hamiltonian up to second-order in perturbation
(even for finite anti-adiabatic values $t/\omega_0 \lesssim1$);
 the small parameter [$t/(g \omega_0)$] of the perturbation will be derived below.                          
In the above equation (\ref{H^2}), unlike in Ref.~\onlinecite{sdadys},
 the term $H_j$ produces phonons
 at sites $j$, $j-1$,
 and $j+1$. Hence, we get non-zero contributions only when the
 index $i = j-2,~ j-1,~ j,
~ j+1$, or $j+2$. Then after some tedious algebra one obtains 
\begin{eqnarray}
&&\!\!\!\!\!\!\!\!\!\!\!
-H^{(2)}
 \frac{\omega_0}{t^2 e^{-6 g^2}}
=
\nonumber \\
&&\!\!\!\!\!\!\!\!\!\!\!
\sum_j \left \{ [ n_j (1-n_{j+1}) + (1-n_j) n_{j+1} ] \right .
\nonumber \\
&&[F_3(4,1,1)+2F_2(4,1)+F_1(4)+ 2F_1(1)+F_2(1,1)]
\nonumber \\
&&+  [c^{\dagger}_{j-1}(1-2n_j) c_{j+1} + {\rm H.c.} ]
[2F_1 (2)+F_2 (2,2)]
\nonumber \\
&&+  2 [c^{\dagger}_{j-2} c_{j-1} c^{\dagger}_{j+1} c_{j} + {\rm H.c.} ]
F_1 (1)
\nonumber \\
&&\left .
 + 2 [c^{\dagger}_{j-1} c_{j-2} c^{\dagger}_{j+1} c_{j} + {\rm H.c.} ]
F_1 (-1)
 \right \} ,
\label{eq:H2}
 \end{eqnarray}                               
where
\begin{eqnarray*}
F_n(\alpha_1, ... , \alpha_n ) \equiv \sum_{m_1=1}^{\infty} 
 ...
\sum_{m_n=1}^{\infty}
 \frac {(\alpha_1 g^2)^{m_1} ... (\alpha_n g^2)^{m_n}}
{m_1! ... m_n!(m_1+ ... + m_n)},
 \end{eqnarray*}                               
which for 
large values of 
$g^2$
becomes 
 $F_n 
 \approx \exp ( g^2 \sum_{i=1}^{n} \alpha_i  )/( g^2 \sum_{i=1}^{n} \alpha_i )$
 for $ \sum_{i=1}^{n} \alpha_i  \ge 1$.
In the above Eq.~\eqref{eq:H2}, the last two terms are a direct consequence of the
cooperative nature of the EPI and are negligible
for large 
 $g^2$.
 More importantly, the relative importance of
the various coefficients is noticeably different from the case where 
no cooperative effect exists (as explained below).
For large $g^2$, the effective polaronic Hamiltonian simplifies to be:
\begin{eqnarray}
H^{C}_{eff} = && 
-\left [ 2 g^2 \omega_0 
 +\frac{t^2}{3 g^2 \omega_0} \right ]
\sum_j n_{j+1} (1-n_{j})
\nonumber \\
&&
 - t e^{-3 g^2} \sum_{j} ( c^{\dagger}_{j} c_{j+1} + {\rm H.c.}) 
\nonumber \\
&&
 - \frac{t^2 e^{-2 g^2} }{4 g^2 \omega_0}
\sum_j [c^{\dagger}_{j-1}(1-2n_j) c_{j+1} + {\rm H.c.} ] .
\label{eq:Hpol}
 \end{eqnarray}                               
Notice that the coefficient of the NN hopping is significantly
smaller than the coefficient of the NNN hopping for
large $g^2$ and not-too-small $t/\omega_0$!
This is a {\em key feature resulting from cooperative effects.}
The above effective Hamiltonian
may be contrasted with the following  Hamiltonian $H_{eff}$
for the case where there is no cooperative EPI
 [i.e., $H_{ep} = -\sqrt{2}g \omega_0 \sum_i
n_i (a^{\dagger}_i + a_i)$]
 [see Ref.~\onlinecite{sdadys} and Fig.~\ref{fig:chain}(a)]:
\begin{eqnarray}
H_{eff} = && 
-2 g^2 \omega_0 \sum_j n_j
 -\frac{t^2}{2 g^2 \omega_0} \sum_j n_{j+1} (1-n_{j})
\nonumber \\
&&
 - t e^{-2 g^2} \sum_{j} ( c^{\dagger}_{j} c_{j+1} + {\rm H.c.}) 
\nonumber \\
&&
 - \frac{t^2 e^{-2 g^2} }{2 g^2 \omega_0}
\sum_j [c^{\dagger}_{j-1}(1-2n_j) c_{j+1} + {\rm H.c.} ] .
\label{eq:Hpolold}
 \end{eqnarray}

We now provide an explanation of the above results. 
In Eq.~\eqref{eq:Hpolold}, the coefficient of the 
$ \sum_j n_{j+1} (1-n_{j})$
 term can be understood as resulting from 
a hopping  process
 where an electron at site $j+1$ hops to a neighboring site $j$ 
and back, but the
lattice has no time to distort (relax) {\em locally}
at site $j$ ($j+1$)
and thus 
yields the second-order perturbation energy $-t^2$/(energy change) (see Fig.~2 of Ref.~\onlinecite{sahinur}).
On the other hand, the coefficient of
$\sum_j [c^{\dagger}_{j-1}(1-2n_j) c_{j+1} + {\rm H.c.} ] $
results when, 
in the intermediate state, 
site $j$ does not distort/relax during
 hopping and thus yields 
$ t\exp[-2 g^2] \times \frac{t}{2 g^2 \omega_0}$ where
$ t\exp[-2 g^2] $ is due to 
time $\left (\frac{\hbar}{te^{-2g^2}} \right )$ taken to distort the site $j+1$ (see Fig.~2 of Ref.~\onlinecite{sahinur}).
In the above non-cooperative case, the NN hopping dominates over
the NNN hopping 
in the small polaron limit.

Using the above logic we see that the
higher order terms in perturbation theory,
for both cooperative and non-cooperative cases,
 are dominated by the process where
an electron hops back and forth between the same two sites. The
dominant term to $k$th order is approximately given for even $k$ by 
\begin{eqnarray}
  \omega_0 \left [ \frac{t}{ g \omega_0} \right  ]^k 
\sum_j n_{j+1} (1-n_{j}) ,
 \end{eqnarray}                               
while for odd $k$  by
\begin{eqnarray}
t e^{-\gamma g^2}
\left [ \frac{t}{ g \omega_0} \right ]^{k-1} 
 \sum_{j} ( c^{\dagger}_{j} c_{j+1} + {\rm H.c.}) , 
 \end{eqnarray}                               
where $\gamma$ is 2 for the non-cooperative case and 3 for the cooperative one.
Since each term in the perturbation theory should be smaller than
$\omega_0$, we see that the small parameter in our perturbation theory
 is $t/(g \omega_0)$ \cite{sdys}.

Here, a few observations are in order.
Firstly, the cooperative effects, unlike in the Holstein
model's case, raise  
the potential of the site next to an occupied site
 and thus make it unfavorable for hopping.
Consequently, in Eq.~\eqref{eq:Hpol} as compared to Eq.~\eqref{eq:Hpolold},
 the exponent is larger for the NN hopping
and also the denominators of the coefficients 
are similarly larger for the hopping-generated
NN interaction
 and for the NNN hopping. Next,
 Lau {\it et al.} \cite{sawatzky}
obtain the same energy expression for {\it a single polaron}
as that given by Eq.~\eqref{eq:Hpol} (when $n_j=0$).
Additionally, in Ref.~\onlinecite{tvr}, the authors explain the ferromagnetic
insulating behavior in low-doped manganites by using the
non-cooperative 
hopping-generated NN interaction [i.e., second term on RHS
of Eq.~\eqref{eq:Hpolold}]
after modifying the hopping term for double-exchange effects.
From Eq.~\eqref{eq:Hpol},
 we see that cooperative phenomenon must be taken into
account as it reduces the ferromagnetism generating interaction
strength by a factor of $1.5$.
Lastly, the authors of Refs. \onlinecite{alex1,alex2} study the formation
of bipolarons using Fr\"ohlich polarons with spin degrees of freedom; although
they use a Lang-Firsov transformation followed by a Schrieffer-Wolff transformation
 (which is similar to our type of perturbation theory), they nevertheless do not consider the 
dominant NNN hopping effects which are central to our treatment.

In the next few sections, we will  analyze the effective
polaronic Hamiltonian given by Eq.~\eqref{eq:Hpol}
and show that there is a period-doubling QPT from a Luttinger
liquid to a conducting CDW
when the coupling $g$ increases at fixed adiabaticity $t/\omega_0$:
the transition is a consequence of enhanced NNN hopping and pronounced NN repulsion.
We employ a modified Lanczos algorithm \cite{gagliano} [and use antiperiodic (periodic)
boundary conditions for even (odd) number of fermions] to 
 study the QPT in the system.
In all our numerical calculations
involving the effective polaronic Hamiltonian, 
we used the series $F_n(\alpha_1, ... , \alpha_n )$ given in Eq. (\ref{eq:H2}) and not the approximate
coefficients in Eq. (\ref{eq:Hpol}).
\section{Density-density correlation function and structure factor}

In this section, to characterize correlations and analyze QPT, we present the relevant
formulae for the density-density correlation function and the  structure factor. 
 The two-point correlation function for density fluctuations of electrons at a  distance $l$ apart is given by
\begin{equation}
 W(l)=\frac{4}{N}\sum_j\left[\langle n_jn_{j+l}\rangle-\langle n_j\rangle\langle n_{j+l}\rangle\right] ,
\label{wl}
\end{equation}
 with filling-fraction (FF) $\langle n_j\rangle =\frac{N_p}{N}$ where $N$ is the 
total number of sites and $N_p$ is the total number of electrons in the system.
  Then the structure factor, which is the Fourier transform of $W(l)$,
is given by
\begin{equation}
 S(k)=\sum_le^{ikl}W(l) ,
\end{equation}
where wavevector $k=\frac{2n\pi}{N}$ with n=1,2,.....,N. Now, we observe that
\begin{equation}
 S(\pi)=\left(\sum_{l_{\rm even}}-\sum_{l_{\rm odd}}\right)W(l) , \nonumber 
\end{equation}
with
\begin{equation}
\sum_{l_{\rm even}}W(l)=\frac{2\langle(\hat N_e-\hat N_o)^2\rangle}{N}  , \nonumber
\end{equation}
and
\begin{equation}
\sum_{l_{\rm odd}}W(l)=-\frac{2\langle(\hat N_e-\hat N_o)^2\rangle}{N}  ,\nonumber 
\end{equation}
where $\hat N_e=\sum_{j_{\rm even}}n_j$  $(\hat N_o=\sum_{j_{\rm odd}}n_j)$ is the number operator which gives 
the total 
number of electrons at even (odd) sites.
Hence, we obtain the simple expression
\begin{equation}
 S(\pi)=\frac{4\langle(\hat N_e-\hat N_o)^2\rangle}{N} .
\label{eq:spi}
\end{equation}

 We will now analyze the situation where only one sub-lattice is occupied
and obtain some exact results. 
When we consider 
 odd values of $l$, we note  that
\begin{equation}
 \langle n_jn_{j+l}\rangle=0. \nonumber
\end{equation}
Hence, from Eq.~\eqref{wl} we get
\begin{equation}
 W(l_{\rm odd})=-\frac{4N^2_p}{N^2}  .
\label{eq:wlodd}
\end{equation}
Next, we observe that the GS becomes an eigenstate of the operators
$\hat N_e$ and $\hat N_o$ with the eigenvalues $N_p$ ($0$)  and $0$ ($N_p$)
respectively if the even-site (odd-site)
sub-lattice is occupied.
Consequently, we get
\begin{equation}
\left[S(\pi)\right] = \left[S(\pi)\right]_{\rm max}=\frac{4N^2_p}{N} ,
\label{eq:spimax}
\end{equation}
where $\left[S(\pi)\right]_{\rm max}$ is the maximum value that  $S(\pi)$ can attain. 

{ To analyze the QPTs, we can treat the rescaled value of $S(\pi)$ as the order parameter
 $S^*(\pi)$ defined as follows:
\begin{equation}
S^*(\pi)= \frac{S(\pi)-\left[S(\pi)\right]_{\rm min}}
{\left[S(\pi)\right]_{\rm max}-\left[S(\pi)\right]_{\rm min}}  ,
\end{equation}
where $\left[S(\pi)\right]_{\rm min}$ is the minimum value of 
$S(\pi)$; 
 consequently, $S^*(\pi)$ varies
from 0 to 1 during the phase transition.
}

In the next section, we will study the limiting case of large EPC values where the NNN hopping 
is the only relevant transport mechanism in the CBM model leading to the t$_2$-V model [see Eq. (\ref{eq:t2v})].

\section{Analysis of the $t_2-V$ model -- a limiting
case of the CBM model}
{ The effective Hamiltonian for the CBM model contains three
terms, namely, NN hopping, NNN hopping, and NN repulsion [as can be seen from Eq.~\eqref{eq:Hpol}].
 There are two possible extreme cases of the CBM model corresponding to small and 
large values of the EPC $g$. For small values of $g$ ($\sim 1$), NN hopping dominates
over NNN hopping; consequently, Eq.~\eqref{eq:Hpol} reduces to
\begin{equation}
H_{tV}
\equiv - t \sum_{j} ( c^{\dagger}_{j} c_{j+1} + {\rm H.c.})
 + V \sum_{j} n_j n_{j+1},
\label{eq:tv}
\end{equation}
 which is the well studied t-V model \cite{gagliano,haldane} with $t/V << 1$ at the small values
of $g$ ($\sim1 $) considered.

On the other hand, for large values of $g$, NNN hopping dominates over NN hopping and
 Eq.~\eqref{eq:Hpol} can be simplified to
\begin{eqnarray}
H_{t_2V} \equiv &&  
 - t_2 \sum_j (c^{\dagger}_{j-1}(1-2n_{j}) c_{j+1} + {\rm H.c.} ) 
\nonumber \\
&&
+ V \sum_j n_{j} n_{j+1} ,
\label{eq:t2v}
 \end{eqnarray}   
 which we shall call as the t$_2$-V model; here, since EPC is large (i.e., $g \gtrsim 3$), $t_2/V << 1$. 
 However, owing to the novelty of the model, we shall study it
[i.e.,  Eq. (\ref{eq:t2v})] for arbitrary values of $t_2/V$
in rings with even number of sites.
Next, for $t_2/V << 1$,
we note that the system  always has alternate sites (i.e., one sub-lattice) occupied
for less than half-filling and above half-filling the other sub-lattice gets
filled. This can be explained, for less than half-filling, as follows.
At large repulsion, we shall compare the energy for the following two situations:
\begin{enumerate}
 \item When there are $m_A>0$ ($m_B>0$) electrons in sub-lattice A (B).
 \item When all the $m_A+m_B=N_p$ electrons are in one sub-lattice only.
\end{enumerate}
In case $1$, each electron in sub-lattice B has $m_{B}-1$ sites blocked in B 
by other electrons in B and at least [at most] $m_{A}+1$ [$2m_{A}$] sites blocked in B by
 electrons in sub-lattice A; one can similarly argue for the electrons in sub-lattice A. Thus in sub-lattice
B(A), each electron can hop to
at most $\frac{N}{2}-m_{A(B)}-m_{B(A)}$ unblocked sites and at least $\frac{N}{2}-2 m_{A(B)} -m_{B(A)}+1$ 
unblocked sites.

In case $2$,  each electron has $m_A+m_B-1$ sites blocked by the other electrons in the same sub-lattice. Hence, each 
electron has exactly $\frac{N}{2}-m_A-m_B+1$ unblocked sites to hop to.
 At large repulsion, since case $2$ gives electrons more number of unblocked sites to hop to,
we see that the total energy
is the lowest when all the electrons are present in the same sub-lattice.

As for the other extreme
situation $V=0$, for even number of electrons,
the model has both sub-lattices equally occupied.

%This is due to the fact that at large repulsion,
%as we fill up the lattice with electrons  one after the other,
% each new electron added to the system has more
%number of sites to hop to on the same sub-lattice occupied
%by the previous electrons.

In the t$_2$-V model, {\it the ground state energy has a 
slope discontinuity, with the energy increasing
up to a critical value, after which it is constant 
for FFs $\frac{1}{4}$, $\frac{1}{3}$, and $\frac{1}{2}$
[as shown in Fig.~\ref{fig:tp_GSESPI} (a)]}.
We will now 
show clearly that as the interaction strength increases, {\it at a critical
value of $V/t_2$,
Ising $Z_2$ symmetry
 (i.e., both sub-lattices being equally populated)
 is broken and only a single sub-lattice
is occupied}. As depicted in Fig.~\ref{fig:tp_GSESPI} (b), 
the structure factor $S(\pi)$ jumps from zero to its maximum
value [given by Eq.~\eqref{eq:spimax} for FFs $\frac{1}{4}$, 
$\frac{1}{3}$, and $\frac{1}{2}$] 
 indicating {\it explosive} first-order QPT from a Luttinger liquid
to a CDW.

\begin{figure}[t]
\includegraphics[height=8.5cm,width=4.5cm,angle=-90]{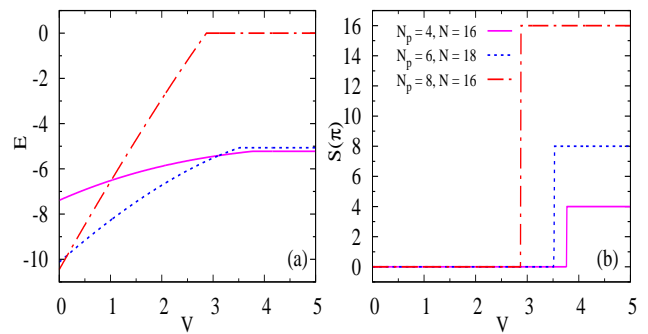}
 \caption{(Color online) Plots, of (a) ground state energy (E) and (b) structure factor value
$S(\pi)$, as a function of interaction strength (V)
for the t$_2$-V model in rings with N sites, Np electrons, and hopping t$_2$ = 1.
}
\label{fig:tp_GSESPI}  
\end{figure}

Next, we observe that the number of electrons in even and odd
sub-lattices are conserved quantities
  for the t$_2$-V model. Therefore, 
GS of the system is an eigenstate of both $\hat N_e$ and $\hat N_o$ with
eigenvalues $N_e$ and $N_o$, respectively. Hence, Eq.~\eqref{eq:spi} simplifies to 
\begin{equation}
 S(\pi)=\frac{4(N_e - N_o)^2}{N} .
\end{equation}
Then, when $Z_2$ symmetry is respected,
 for even number of electrons $N_p = 2N_e = 2N_o$,
we have $S(\pi)=0$ and for odd 
value of $N_p$ we have $S(\pi)= 4/N$. We find that 
 at a critical interaction strength,
as shown in  Fig.~\ref{fig:tp_wlsk}, the following dramatic changes occur:
(i) the structure factor $S(\pi)$ 
jumps 
 from 
 $0$ to 
its maximum value $4 N_p^2/N$; 
(ii) $W(l {\rm odd})$ also jumps 
 to its large $V$ value of $-\frac{4N^2_p}{N^2}$;
and (iii)  $W(l {\rm even})$ (for $l \neq 0$) too jumps and its
final value at half-filling is 1.
For a fixed $t_2$ and $N$, the critical value $V^N_C$ of $V$
 increases monotonically
as $N_p$ decreases. For $N=16$, $N_p = 2$, and $t_2=1$,
 we get $V^{16}_{C} \approx
4$. From finite size scaling for half-filling, using 
$V^N_{C} - V^{\infty}_C \propto 1/N^2$ and system size $N \le 20$,
we obtain $V^{\infty}_C \approx 2.83$.

We see from the above analysis that, at a critical repulsion,
 the system undergoes a discontinuous
transition to a {\em conducting commensurate CDW} state away from half-filling
while at half-filling one obtains a Mott insulator.
Usually commensurate CDW's are insulating (see Ref. \onlinecite{Gruner})
 whereas our model surprisingly predicts a conducting commensurate CDW.
Furthermore, quite unlike the Peierls transition,
 the {\it period of the CDW is  independent of density}!

\begin{figure}[b]
\centering
\includegraphics[height=8.5cm,width=8.5cm,angle=-90]{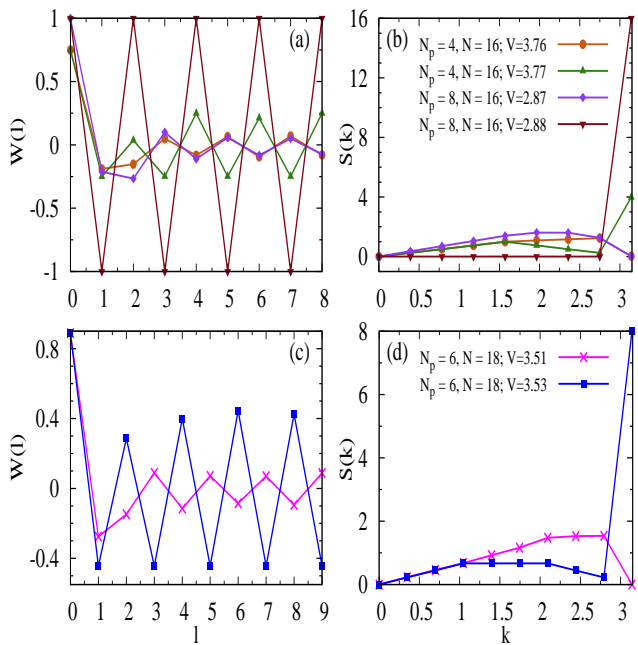}
\caption{(Color online) Plots of the density-density correlation function W(l)
and structure factor S(k) for the t$_2$-V model on a N-site ring with Np electrons,
interaction strength V, and hopping t$_2$ = 1.
{Structure factor S(k) in (b) and (d) correspond to plots of W(l) in (a) and (c),
respectively.}
}
\label{fig:tp_wlsk} 
\end{figure}
}
\section{Analysis of the CBM model}
To analyze the QPT at various FFs of a system governed by
the effective
polaronic Hamiltonian given in Eq.~\eqref{eq:Hpol},
 we performed our calculations at 
values of the adiabaticity $t/\omega_0 < 1$ and $g \ge 1$ such that the small parameter
$t/(g\omega_0) < 1$ and $t e^{-3 g^2} \ll \omega_0$. Here we report only for the
 conservative case $t/\omega_0 = 0.1$
since the results at other values of $t/\omega_0 < 1$ are qualitatively similar 
(as shown in appendix \ref{app:tw}).

As the value of $g$ increases (in the regime of study $1 \le g \le 3.5$), NN
and NNN hoppings compete and
the system gradually transits from a large-V ~t-V model to a large-V ~t$_{2}$-V model; thus, at
values of $g \sim 1$ we expect the system to be a Luttinger liquid while at $g \sim 3$ we should get a CDW.
In the next sub-sections we will demonstrate that the system indeed undergoes a Luttinger
liquid to a {\it conducting} CDW transition with the QPT being second-order in nature.
The QPT discussed in this work is quite different from
the metallic Luttinger liquid to {\it insulating} CDW transition studied by many authors \cite{poilblanc,ortolani,hohenadler}
 in a system
with only NN hopping and long-range Coulomb interaction.

\subsection{Study of density-density correlation function, structure factor, and order parameter}
\begin{figure}[t]
\centering
\includegraphics[height=8.5cm,width=8.5cm,angle=-90]{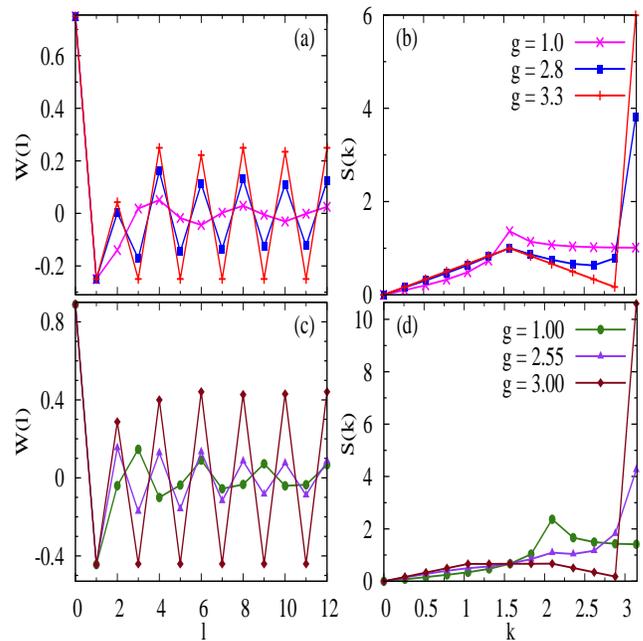}
\caption{(Color online) Density-density correlation function $W(l)$
in the CBM model at
 $\frac{t}{\omega_0}=0.1$ and $N=24$ for
 $(a)$ $\frac{1}{4}$-filling; 
and  $(c)$ $\frac{1}{3}$-filling.
Structure factor $S(k)$ at 
$(b)$ $\frac{1}{4}$-filling;
and $(d)$ $\frac{1}{3}$-filling corresponding to plots of $W(l)$ in $(a)$ and $(c)$ respectively. }
\label{fig:wlsk} 
\end{figure}

First, we calculated $W(l)$ and $S(k)$ at FFs $\frac{1}{4}$ and $\frac{1}{3}$ numerically
and the results are displayed in Fig.~\ref{fig:wlsk}. 
 Upon tuning the EPC $g$, 
 the density-density correlation function $W(l)$ gradually changes its nature
from decaying to oscillatory thereby exhibiting long range order;
 it then attains the value given
by Eq.~\eqref{eq:wlodd} at all odd values of $l$ corresponding to the state of only
one sub-lattice being occupied 
[see Figs.~\ref{fig:wlsk}(a) and (c)]. Furthermore, the structure factor value $S(\pi)$ increases
 upon increasing $g$ 
and attains the maximum value given by Eq.~\eqref{eq:spimax} [see Figs.~\ref{fig:wlsk} (b) and (d)].
 These observations assert
that the system undergoes QPT from a Luttinger liquid to a conducting commensurate CDW
 state away from half-filling with period-doubling.
Thus, at a critical value of $g$,
the Ising $Z_2$ symmetry (i.e., both sub-lattices being equally populated) is broken.
Quite surprisingly, our model 
 predicts a conducting commensurate CDW without an excitation gap.
 Furthermore, similar to the t$_2$-V model, here too the period of CDW is independent of density 
and is, in fact, twice the 
lattice constant.

\begin{figure}[b]
\includegraphics[height=8.5cm,width=8cm,angle=-90]{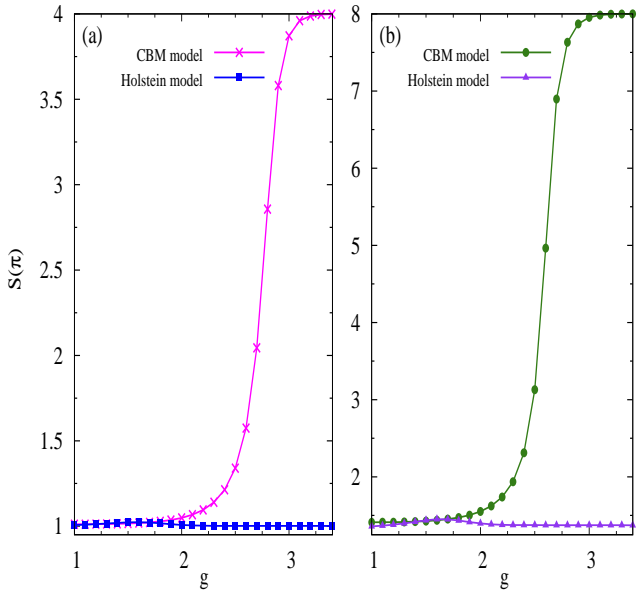}
\caption{(Color online) Structure factor value $S(\pi)$ at $\frac{t}{\omega_0}=0.1$ for 
$(a)$ $\frac{1}{4}$-filling and $N=16$; 
and $(b)$ $\frac{1}{3}$-filling and $N=18$ in our CBM model and
the Holstein model.}
\label{fig:Hols_vs_Coop}   
\end{figure}

 We will now compare $S(\pi)$ versus $g$ behavior 
manifested by our model and the Holstein model in Fig.~\ref{fig:Hols_vs_Coop}.
We see that, while our CBM model appears to undergo a QPT, the Holstein model does not
seem to do so.
We observe that the coefficient
 of NNN hopping for the Holstein model 
(see Eq.~\eqref{eq:Hpolold}) becomes much smaller than that of 
the NN hopping as EPC $g$ increases. Hence, 
the Holstein model, for sufficiently larger values of $g$, behaves like the t-V model;
whereas our model can be approximated by the t$_2$-V model at large $g$.
 We know that the t-V model does not
undergo a QPT away from half-filling \cite{haldane}. Therefore,
 the  Holstein model too will not undergo a QPT at a 
 non-half FF (which is consistent with the results of Ref.~\onlinecite{sdadys}). 
Thus, the $Z_2$ symmetry breaking QPT (at non-half filling) in our model is a unique feature
which has no analog in either the Holstein model or the t-V model.
 Furthermore, at half-filling, the t-V model undergoes a QPT when $V=2t$ \cite{gagliano,haldane} while
the Holstein model suffers a QPT at $g > 1$  \cite{sdadys,hamer};
on the other hand our CBM model, for the range of EPC $g$ considered (i.e., $g \ge 1$),
is always deep inside the CDW phase since the coefficient of NN repulsion is much larger 
than the hopping terms.

\begin{figure}[t]
\includegraphics[height=9cm,width=8cm,angle=-90]{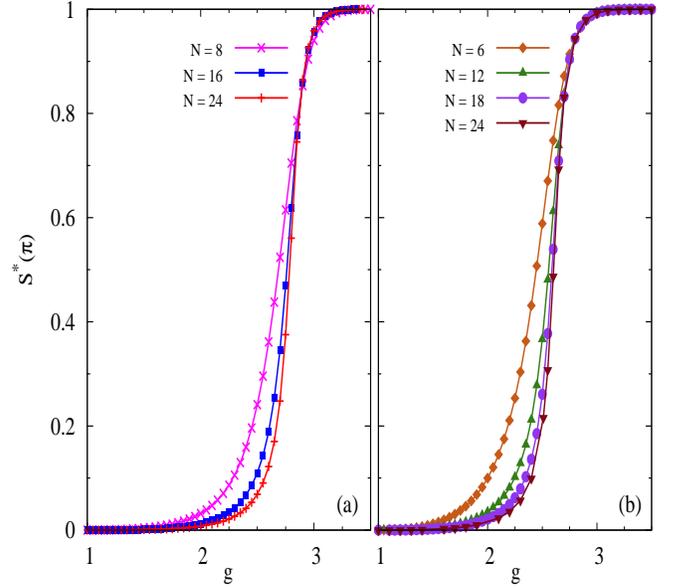}
\caption{(Color online) Order parameter $S^*(\pi)$ in the CBM model at $\frac{t}{\omega_0}=0.1$ for 
 $(a)$ $\frac{1}{4}$-filling;
 and  $(b)$ $\frac{1}{3}$-filling.}
\label{fig:order_parameter}   
\end{figure}

 Plots of the order parameter $S^*(\pi)$
displayed in Fig.~\ref{fig:order_parameter} also reveal signatures of
 QPT  at FFs $\frac{1}{4}$ and $\frac{1}{3}$
 and at different system 
sizes. Moreover, we  observe that the increase in $S^*(\pi)$ becomes sharper as the
system size increases. From the figures it appears that there is either a continuous or weakly first-order 
QPT for both the FFs.  

\subsection{Ground State Fidelity; Fidelity Susceptibility and its Scaling Behavior}
 Although the order parameter $S^*(\pi)$ depicts a QPT,
 but the nature of the transition (whether it is first-order, second-order, or KT-like) 
is not clear. Therefore, we take recourse to the study of 
the ground state fidelity (GSF) and FS to characterize the nature of the QPT. 
  The GSF is defined as the overlap between GSs at 
two different but near values of the control parameter 
(say $g$ and $g+\delta$) as follows:
\begin{equation}
 F(g,\delta)=|\langle\Psi_0(g)|\Psi_0(g+\delta)\rangle| , 
\label{eq:fidelity}
\end{equation}
where $|\Psi_0 \rangle$ is the GS of the system and $\delta$ is a small quantity \cite{zanardi}. It is clear
 from Eq.~\eqref{eq:fidelity} that $F(g,\delta)$ depends on $\delta$.
 On the other hand, the FS \cite{you}, defined below as the second derivative of 
GSF \cite{zanardi},
\begin{equation}
 \chi_F(g) \equiv 
\partial^2_\delta F(g,\delta)|_{\delta=0}=2\lim_{\delta\rightarrow0}\frac{1-F(g,\delta)}{\delta^2} ,
\label{eq:susceptibility}
\end{equation}
 is independent of $\delta$.

 The GS of the system, after transition, becomes two-fold degenerate as 
either of the two sub-lattices, namely even and odd, can have the larger occupancy.
 Now,  any linear superposition of the two degenerate states is also a GS.
 Therefore, the calculated GSF [i.e., the absolute value of the overlap of GS $|\Psi_0 \rangle$ at 
two close by values of the control parameter ($g$ and $g+\delta$)] becomes arbitrary. 
To eliminate arbitrariness in the estimate of GSF, we start with $\Psi_0(g)$ as our initial guess
in the modified Lanczos algorithm to get the GS $\Psi_0(g+\delta)$.

 Next, we point out a mapping that will enable us to perform fidelity calculations
in systems with  sizes larger than the usual sizes accessible to the modified Lanczos technique.
 At $\frac{N_p}{N}$-filling  in our CBM model, when 
 NN repulsion is much larger than both the
 NN and the NNN hoppings,
 our model can be reduced to the following  model 
at $\frac{N_p}{N-N_p}$-filling but without 
NN repulsion 
(for similar analyses, see the treatment of the t-V model in Ref.~\onlinecite{dias} 
and the mapping of the t-V$_1$-V$_2$ model in Ref.~\onlinecite{sahinur_arxiv}):
\begin{eqnarray}
H^{RC}_{eff} &= & 
 - t e^{-3 g^2} \sum_{j} ( c^{\dagger}_{j} c_{j+1} + {\rm H.c.}) 
\nonumber \\
&&
 - \frac{t^2 e^{-2 g^2} }{4 g^2 \omega_0}
\sum_j [c^{\dagger}_{j-1}(1-n_j) c_{j+1} + {\rm H.c.} ] .
\label{eq:Hpol_reduced}
 \end{eqnarray}   
In the NNN hopping term, because of large NN repulsion, 
we have ignored the contribution of the sequential hopping depicted in Fig. 2(c) of
 Ref.~\onlinecite{sahinur}. 
The above prescription reduces the dimension of the Hilbert space
 significantly from $^{N}C_{N_p}$ to $^{N-N_p}C_{N_p}$.
 From Eq.~\eqref{eq:Hpol_reduced}, we also observe that the
new effective Hamiltonian contains
only kinetic terms. Hence, the GS in the CDW phase has to be conducting away from half-filling. 

\begin{figure}[t]
\includegraphics[height=8.5cm,width=8cm,angle=-90]{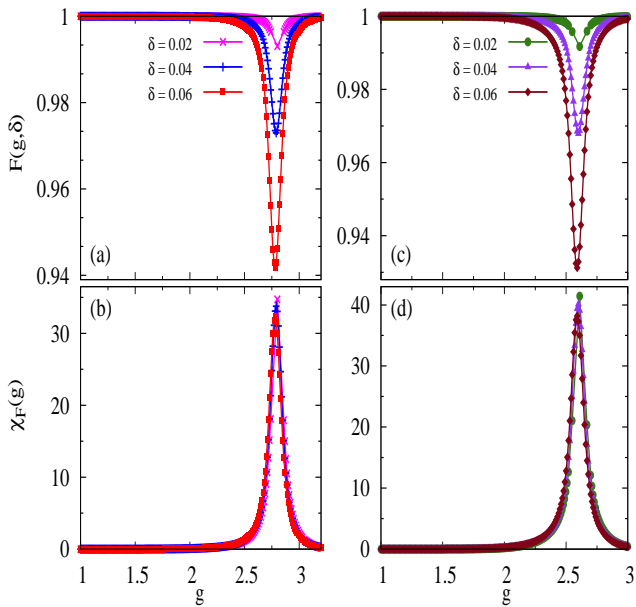}
\caption{(Color online) GSF $F(g, \delta)$ in the CBM model at $\frac{t}{\omega_0}=0.1$ for 
$(a)$ $\frac{1}{4}$-filling and $N=32$;
and $(c)$ $\frac{1}{3}$-filling and $N=30$.
FS $\chi_F(g)$ for $(b)$ $\frac{1}{4}$-filling; and
 $(d)$ $\frac{1}{3}$-filling correspond to the GSF-plots in $(a)$ and $(c)$
respectively.
For the sake of clarity, only selected points are shown for $\delta = 0.02$.}
\label{fig:fid_deltaG}   
\end{figure}

In Fig.~\ref{fig:fid_deltaG}, we depict $F(g,\delta)$ and $\chi_F(g)$ as a function
of $g$ at FFs $\frac{1}{4}$ and $\frac{1}{3}$.
The dip in $F(g,\delta)$ at the critical point increases with the increase in $\delta$. 
This happens because of the fact that the distance
between two ground states in parameter space increases with the increase in $\delta$.
However, $\chi_F(g)$ for different small values of $\delta$ coincide as $\chi_F(g)$ is 
independent of $\delta$ [as can be seen from Eq.~\eqref{eq:susceptibility}]. 

\begin{figure}[b]

\includegraphics[height=8cm,width=8cm,angle=-90]{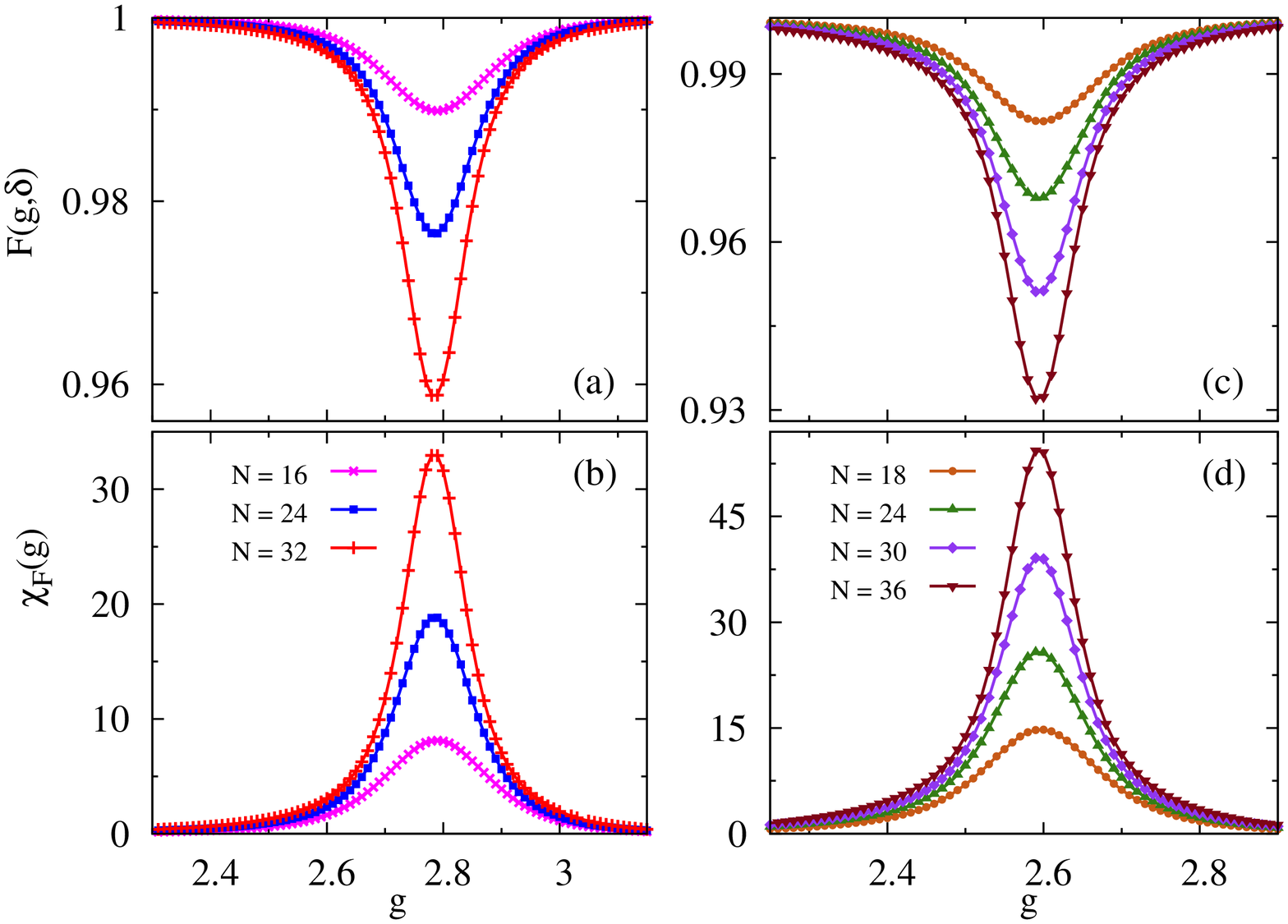} \\
\hspace{.5cm}
\includegraphics[height=8cm,width=3cm,angle=-90]{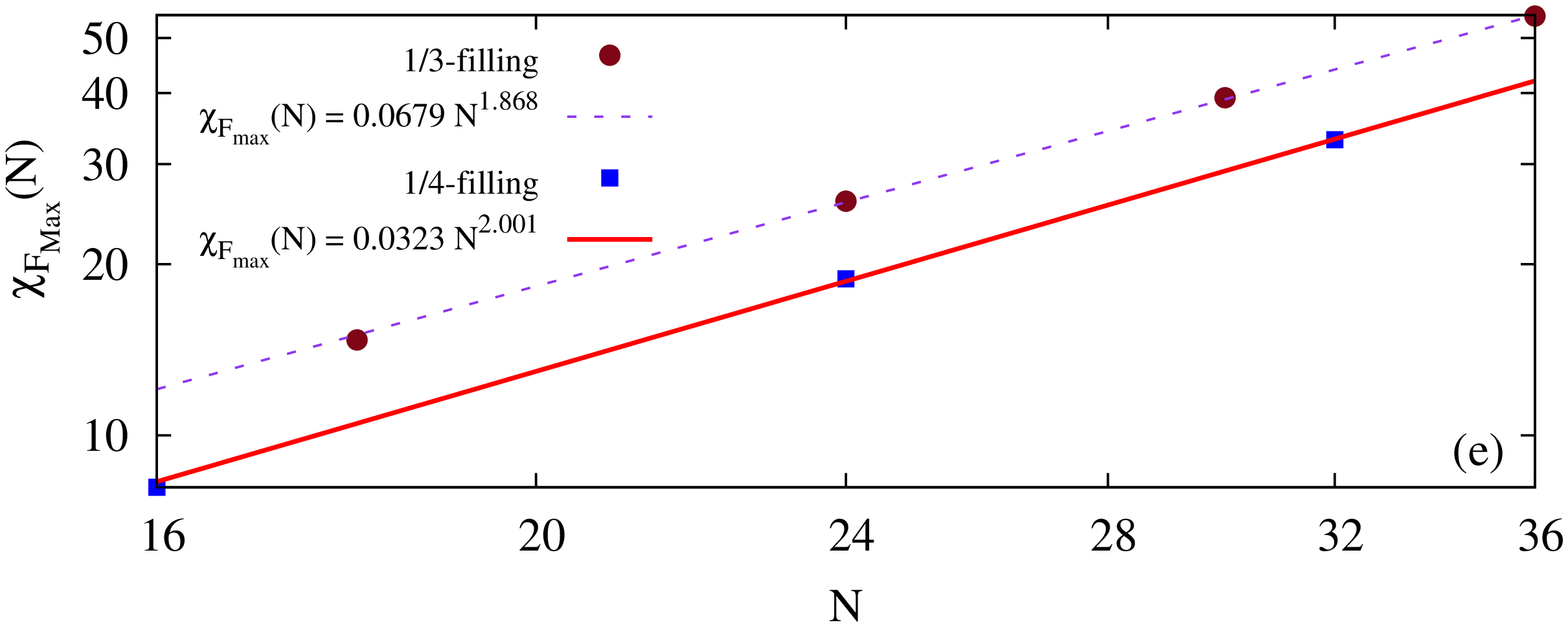}
 \caption{(Color online) GSF $F(g, \delta)$ in the CBM model at $\frac{t}{\omega_0}=0.1$ and $\delta = 0.05$
for $(a)$ $\frac{1}{4}$-filling; and $(c)$ $\frac{1}{3}$-filling. FS $\chi_F(g)$
 for $(b)$ $\frac{1}{4}$-filling; and 
 $(d)$ $\frac{1}{3}$-filling correspond to the GSF-plots in $(a)$ and $(c)$. (e) Plot of the
peak values of FS $\chi_{F_{\rm max}}$(N) versus $N$, on a logarithmic scale, at 
$\frac{1}{4}$-filling and $\frac{1}{3}$-filling and the corresponding power-law fits.}
\label{fig:fid_sys_size}  
\end{figure}

 Additionally, Fig.~\ref{fig:fid_sys_size}
 shows $F(g,\delta =0.05)$, $\chi_F(g)$ and $\chi_{F_{\rm max}}$ (or the
peak FS) for different
system sizes at FFs $\frac{1}{4}$ and $\frac{1}{3}$. The dip (peak) in $F(g,\delta)$ $[\chi_F(g)]$ at 
the extremum point  increases with the system size $N$.
 Furthermore, for a finite system, $\chi_{F_{ \rm max}}$ scales like \cite{gu,gu_review}
\begin{equation}
 \chi_{F_{ \rm max}}\propto N^{\mu}.
\end{equation}
The logarithmic scale plot of the peak FS
value $\chi_{F_{ \rm max}}(N)$ with $N$ shows a linear behavior (see Fig.~\ref{fig:fid_sys_size}$(e)$)
which  confirms a power law divergence of $\chi_{F_{ \rm max}}(N)$ at the extremum point $g_{\rm max}$. 
At large $N$, we obtain $\chi_{F_{ \rm max}}(N)\sim N^{2.001}$
at $\frac{1}{4}$-filling; whereas at  $\frac{1}{3}$-filling we get $\chi_{F_{ \rm max}}(N)\sim N^{1.868}$.
The superextensive power law divergence of $\chi_{F_{\rm max}}$ along
with the dynamical critical exponent value $z \sim 1$ rule out a KT-like transition
 (see appendix \ref{app:sus} for details). 

\begin{figure}[t]
\includegraphics[height=8cm,width=4.5cm,angle=-90]{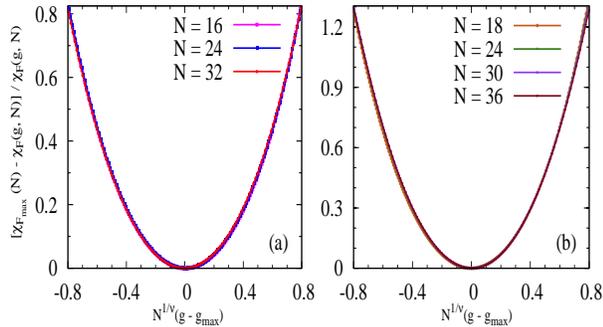}
 \caption{(Color online) Scaling behavior of FS $\chi_F(g, N)$ in the CBM model at $\frac{t}{\omega_0}=0.1$ for
 $(a)$ $\frac{1}{4}$-filling
 yielding $\nu = 1.33\pm 0.01$ and for $(b)$ $\frac{1}{3}$-filling producing $\nu = 1.41\pm 0.01$.}
\label{fig:sus_scale}  
\end{figure}

 In order to examine the possibility of a second-order QPT, 
we consider the following scaling relation \cite{gu,gu_review} for $\chi_F(g)$:
\begin{equation}
 \frac{(\chi_{F_{ \rm max}}(N)-\chi_F(g,N))}{\chi_F(g,N)}=f[N^{\frac{1}{\nu}}(g-g_{\rm max})],
\label{eq:scale} 
\end{equation}
where $\nu$ is the critical exponent of the correlation length.
 Interestingly, a plot of
$[\chi_{F_{ \rm max}}(N)-\chi_F(g,N)]/{\chi_F(g,N)}$ versus $N^{\frac{1}{\nu}}(g-g_{\rm max})$, as depicted in
Fig.~\ref{fig:sus_scale}, shows a nice scaling relation of $\chi_F(g,N)$ with
 $\nu$ taking the values
 $1.33\pm 0.01$ and $1.41\pm 0.01$
 for the best fits to the universal curves
at FFs $\frac{1}{4}$ and $\frac{1}{3}$ respectively. 
The superextensive power law divergence and the scaling behavior
 of $\chi_F$ demonstrate that the QPT is second-order in nature.

Furthermore, as pointed out in Refs.~\onlinecite{gu,gu_review}, average FS $\chi_F(g)/N$
around the critical point $g_c$ scales like
\begin{equation}
 \frac{\chi_F(g)}{N}\propto \frac{1}{\arrowvert g_c-g\arrowvert^\alpha},
\end{equation}
in the thermodynamic limit, with $\alpha$ being a critical exponent.
The three exponents $\alpha$, $\mu$, and $\nu$ are related as\cite{gu,gu_review}
\begin{equation}
 \alpha= \nu(\mu-1).
\label{eq:exponents}
\end{equation}
The values of the critical exponent $\alpha$, on using Eq.~\eqref{eq:exponents}, turn
 out to be $\alpha\simeq 1.33$ and 
$\alpha\simeq 1.22$ for FFs $\frac{1}{4}$ and $\frac{1}{3}$ respectively. On using finite size
  scaling, we find  the critical point $g_c$ values to be $2.785$ and $2.594$ for 
FFs $\frac{1}{4}$ and $\frac{1}{3}$ respectively [based on  positions of 
dips (peaks) of GSF (FS) in Fig.~\ref{fig:fid_sys_size}]. 

\section{Conclusions}
 We derived an effective Hamiltonian for molecular chains involving CBM
at strong EPI. The spinless fermion model considered here
should be relevant to perovskite systems with large onsite coulomb repulsion.
Our analysis shows that our system has an effective Hamiltonian
of the form
\begin{eqnarray}
H_{t-t_2-V}
= &&- t \sum_{j} ( c^{\dagger}_{j} c_{j+1} + {\rm H.c.})
\nonumber \\
&&- t_2 \sum_j (c^{\dagger}_{j-1}(1-2n_{j}) c_{j+1} + {\rm H.c.} ) 
\nonumber \\
&& + V \sum_{j} n_j n_{j+1} ,
\label{eq:tt2v}
\end{eqnarray}
with $t_2 << t$ for small $g$ ($\sim 1$), whereas $t_2 >> t$ for large $g$ ($\gtrsim 3$);
furthermore $V$ is significantly larger than both $t$ and $t_2$ for
all values of EPC ($1 \le g \le 3.5$) considered.
{ Thus, NN and NNN hoppings compete and
the system transits from a large-V ~t-V model (with a Luttinger liquid GS)
to a large-V ~t$_2$-V model (with a period-doubling CDW GS) as $g$ increases;}
our fidelity analysis shows that
the QPT is second-order in nature.
 In the past, a density independent charge ordering
has  indeed been observed in manganite systems (see Fig.~2 in Ref.~\onlinecite{Littlewood}).
However, since the dimensionality and number of bands are different, our
findings are not directly related to these reported results.
Although the reported calculations were performed for a conservative value of the 
adiabaticity $t/\omega_0 = 0.1$, we find that our results are qualitatively similar
in the whole anti-adiabatic regime of $t/\omega_0 < 1$ (as shown in appendix A).
Furthermore, we provide one more model system where the utility of GSF and FS in studying the nature of
 QPT is clearly demonstrated.

\section{Acknowledgments}
One of the authors (S. Y.) would like to thank P. B. Littlewood, S. Kos, 
 D. E. Khmelnitskii,  T. V. Ramakrishnan,  Diptiman Sen, and M. Q. Lone for valuable discussions and KITP for hospitality.

\appendix
\section{}
\begin{figure}[t]
\includegraphics[height=5.9cm,width=5.5cm,angle=-90]{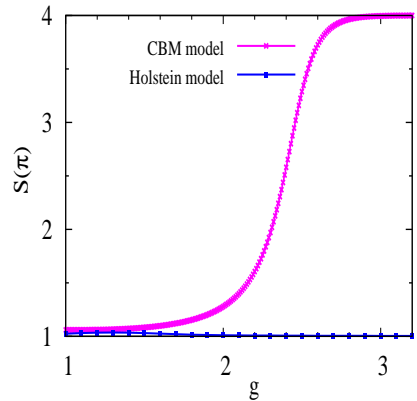}
\caption{(Color online)  Plots of structure factor value $S(\pi)$  at $\frac{1}{4}$-filling for $N=16$ and $\frac{t}{\omega_0}=0.5$.}
\label{fig:strtw05}
 \end{figure}

In this appendix, we study our CBM model at different values of the adiabaticity 
parameter $\frac{t}{\omega_0}$ and for
 system size $N=16$ and FF  $\frac{1}{4}$. 
Firstly, we wish to point out that, in the entire anti-adiabatic 
regime, the $Z_2$ symmetry breaking captured by $S(\pi)$ is a 
novel feature of our CBM model which is not present in the Holstein model;
this claim is endorsed by Figs.~\ref{fig:Hols_vs_Coop} and \ref{fig:strtw05}. 

\begin{figure}[b]
\includegraphics[height=6cm,width=5.5cm,angle=-90]{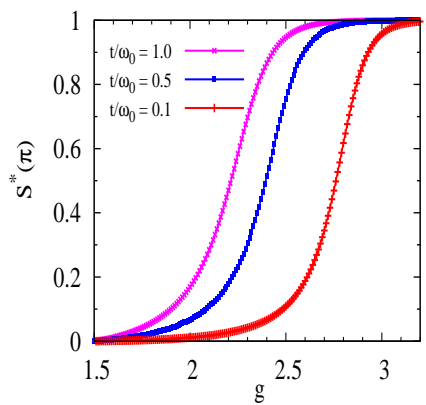}
\caption{(Color online) Order parameter $S^*(\pi)$ in the CBM model at $\frac{1}{4}$-filling, $N=16$, and
for different 
values of $\frac{t}{\omega_0}$.}
\label{fig:strtw}
 \end{figure}

\begin{figure}[t]
\includegraphics[height=8cm,width=4.5cm,angle=-90]{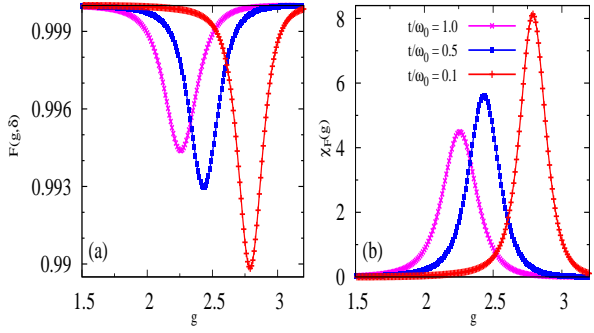}
\caption{(Color online) Plots of $(a)$ GSF $F(g,\delta)$ at $\delta=0.05$ and $(b)$ 
corresponding FS in the CBM model at different values of $\frac{t}{\omega_0}$ and for 
$N=16$ and $\frac{1}{4}$-filling.} 
\label{fig:fidtw}
\end{figure}

As depicted 
in Fig.~\ref{fig:strtw}, for various values of $\frac{t}{\omega_0}$,
the order parameter $S^*(g)$ 
rises from $0$ to $1$ when the  control parameter $g$ is increased;
larger values of $\frac{t}{\omega_0}$ lead to QPT occurring at smaller values of $g$.
\label{app:tw}
The GSF $F(g,\delta)$ and the corresponding FS, at different values of $\frac{t}{\omega_0}$, are 
portrayed in Figs.~\ref{fig:fidtw}(a) and \ref{fig:fidtw}(b) respectively. The dip (peak) in the
GSF (FS) occurs at smaller 
values of $g$ when the adiabaticity $\frac{t}{\omega_0}$ assumes larger values.

From the above analysis, we find that our model exhibits qualitatively similar behavior 
for all values of the adiabaticity parameter $\frac{t}{\omega_0}\leq 1$. Furthermore,
we expect similar behavior at other FFs and system sizes as well.

\section{}
\label{app:sus}
\begin{figure}[t]
\vspace{.5cm}
\includegraphics[height=8cm,width=3.5cm,angle=-90]{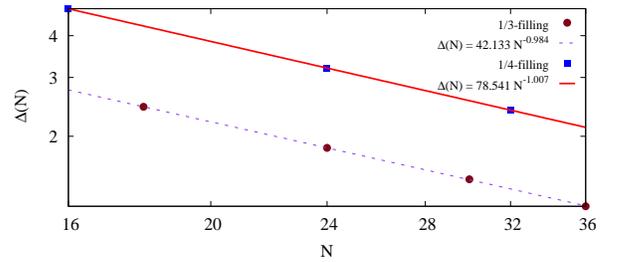}
\caption{(Color online) A log-log plot of the excitation energy $\Delta (N)$ versus $N$
in the CBM model at 
$\frac{1}{4}$-filling and $\frac{1}{3}$-filling and the corresponding power-law fits. }
\label{fig:energy_gap}
 \end{figure}

In this appendix, we will describe a new general approach to identify a KT transition. 
The FS can also be written \cite{you} as
\begin{equation}
\chi_F(g)=\sum_{n\neq0}\frac{|\langle\Psi_n(g)|H_I|\Psi_0(g)\rangle|^2}{\left[E_n(g)-E_0(g)\right]^2} ,
\label{eq:expan_sus}
\end{equation}
where $H_I$ is the QPT driving Hamiltonian.
 On the other hand, according to perturbation theory, the second order perturbation to the GS energy
takes the form
\begin{equation}
 E_0^{(2)}(g)=\sum_{n\neq0}\frac{|\langle\Psi_n(g)|H_I|\Psi_0(g)\rangle|^2}{\left[E_0(g)-E_n(g)\right]} .
\label{eq:energy} 
\end{equation}
At the KT-transition point, both the numerator  and denominator
of Eq.~\eqref{eq:energy} tend to zero with system size in
exactly the same manner; hence no divergence results.
At the extremum point, the mass gap typically vanishes with system size as
\begin{equation}
 \Delta\equiv E_1(g_{\rm max})-E_0(g_{\rm max})\sim \frac{1}{N^z},
\label{eq:excitation}
\end{equation}
where $z$ is the dynamical critical exponent.
 Thus, for a KT-transition the FS at the extremum point  exhibits the following behavior 
\begin{equation}
 \chi_{F_{\rm max}}\sim N^z.
\end{equation}
However, (in contrast to the KT-transition) for a first-order or a second-order transition, the 
numerator of Eq.~\eqref{eq:energy} (at the extremum point) does not tend to 
zero as fast as the denominator and hence divergence occurs in the thermodynamic limit.
 Therefore, the divergence in Eq.~\eqref{eq:energy} leads to an even
stronger power law divergence in the FS at the extremum point 
(as can be seen from  Eq.~\eqref{eq:expan_sus}):
\begin{equation}
\chi_{F_{\rm max}}\sim N^\mu, 
\end{equation}
where $\mu>z$.
Now, in Fig.~\ref{fig:energy_gap}, we depict 
variation of the mass gap $\Delta(N)$ with $N$
 on a logarithmic scale and observe linear behavior. At large $N$, we obtain 
$\Delta(N)\sim \frac{1}{N^{1.007}}~({\rm i.e.,}~z=1.007)$ at $\frac{1}{4}$-filling
while at $\frac{1}{3}$-filling we get $\Delta(N)\sim \frac{1}{N^{0.984}}~({\rm i.e.,}~z=0.984)$;
whereas from Fig.~\ref{fig:fid_sys_size}$(e)$, we find $\chi_{F_{ \rm max}}(N)\sim N^{2.001}~({\rm i.e.,}~\mu=2.001)$
at $\frac{1}{4}$-filling while at  $\frac{1}{3}$-filling we get $\chi_{F_{ \rm max}}(N)\sim N^{1.868}~({\rm i.e.,}~\mu=1.868)$.
Clearly, $\mu>z$ for both the FFs which rules out the possibility of a KT-transition.
 Hence, the QPT in our model is either first-order or second-order. 

Lastly, we would like to mention that another approach, 
 involving examining the global geometric entanglement, to detect the elusive KT quantum phase transition has
been recently reported\cite{wei,alet}.

\end{document}